\pdfoutput=1
\documentclass[prl,twocolumn,showpacs,preprintnumbers,amsmath,nofootinbib,superscriptaddress,amssymb]{revtex4}

\usepackage{graphicx}
\usepackage{dcolumn}
\usepackage{bm}
\usepackage{eucal}
\usepackage{color}
\usepackage{verbatim}

\begin{document}

\preprint{APS/123-QED}

\title{Random Matrix approach to collective behavior and bulk universality\\ in protein dynamics}

\author{Raffaello Potestio}

\affiliation{SISSA - Scuola Internazionale Superiore di Studi Avanzati, Via Beirut 2/4 - 34151 Trieste, Italy}

\author{Fabio Caccioli}

\affiliation{SISSA - Scuola Internazionale Superiore di Studi Avanzati, Via Beirut 2/4 - 34151 Trieste, Italy}
\affiliation{Istituto Nazionale di Fisica Nucleare, sezione di Trieste, Italy}

\author{Pierpaolo Vivo}
\affiliation{ICTP - Abdus Salam International Centre for Theoretical Physics, Strada Costiera 11 - 34151 Trieste, Italy}



\date{\today}

\begin{abstract}

Covariance matrices of amino acid displacements, commonly used to characterize the large-scale movements of proteins, are investigated 
through the prism of Random Matrix Theory. Bulk universality is detected in the local spacing statistics of noise-dressed eigenmodes,
which is well described by a Brody distribution with parameter $\beta\simeq 0.8$. This finding, supported by other consistent indicators, implies
a novel quantitative criterion to single out the collective degrees of freedom of the protein from the majority of high-energy, localized vibrations.

\end{abstract}


\maketitle

{\it Introduction.} 
Proteins are biomolecules of essential importance for biological aspects ranging from  structural (e.g. viral capsids, cytoskeleton) to biochemical ones (e.g. enzymes).
The biological functionality of a protein often relies on its capability to undergo large-scale conformational changes \citep{perutz,Zen:2008:Protein-Sci:18369194}. In order to characterize such motions, a convenient starting point is the Displacement Covariance Matrix (\textsf{DCM})
$\mathcal{C}_{i j} = \langle \delta r_i \delta r_j \rangle$, where $\delta r$ is the deviation of a protein's backbone ($\mathrm{C}_\alpha$) atom from a given reference structure (in this notation $i, j$ indicate both amino acid and Cartesian indices) and $\langle \cdots \rangle$ indicates the ensemble average.

The eigenmodes of the \textsf{DCM} with largest eigenvalues, corresponding to low-energy excitations, are responsible for large-scale, collective motions of the protein. These modes play a major role in the molecule's functionality, and a handful of them account for a large fraction of the overall fluctuation \citep{Noguti:1982:Nature:7070518}.
The other more numerous eigenmodes instead describe high-energy, localized vibrations which are more heavily affected by the fine details of structure and interactions.

Atomistic force-field Molecular Dynamics (\textsf{MD}) is a widely employed tool to investigate the dynamical properties of a protein. The duration of a \textsf{MD} trajectory, however, 
poses a problem in the sampling of the phase space \citep{garcia}, since the essential space can vary from run to run. Nevertheless, 
the principal components are more robust than the high-energy sector of the spectrum and are consistent amongst independent simulations.
It is therefore highly desirable to single out the statistically significant low-energy modes from the bulk of `noise-dressed' ones. Far-reaching implications include the description of large-scale fluctuations in terms of a suitable set of collective coordinates \cite{Pontiggia:2008:bj} and the identification of a few relevant modes as a basis for coarse-grained models of structure and dynamics \cite{potestio2009}.

The purpose of this Letter is to address this issue quantitatively. We propose a novel criterion for statistical significance based on the comparison with \emph{local} eigenvalue statistics from Random Matrix Theory (\textsf{RMT}). Applications of \textsf{RMT} to biophysical issues are so far rather limited 
\cite{ciliberti2006,luo2006,lacelle1984,majumdar2005,sener2002,bandyopadhyay2007,zee2005}.
Our new approach relies on an \emph{ensemble} of \textsf{DCM}, rather than a single instance.
This permits to characterize in a statistical sense the low-energy subspace of a specific protein against a `null' model with random correlations.
The sharp onset of \textsf{RMT}-like level spacing statistics 
from the first few modes onwards, along with other consistent indicators, signals a clear-cut separation of the spectrum in two statistically incompatible regions, associated with discordant dynamical properties.


{\it Preliminaries.}
In order to collect a \textsf{DCM} ensemble we resort to exactly-solvable Elastic Networks Models (\textsf{ENM})
\cite{hinsen98}, rather than other available strategies such as the calculation of the Hessian matrix \citep{brooks}.
Indeed it is known \cite{tirion} that the low-energy modes of the spectrum, the most interesting
for the present work, are well reproduced by \textsf{ENM} with lighter computational effort.


Our matrix ensemble is obtained from a \textsf{MD} simulation carried out in \cite{Pontiggia:2008:bj} on E. Coli
Adenylate Kinase, a $N=214$ amino acid single-chain phosphotransferase protein \cite{Lou_J-Phys-Chem-B_2006_2}.
During this 50 ns simulation, the molecule explores many different free energy minima, or \textit{substates}. Within each substate, the protein fluctuates around a well-defined average structure.
The 4000 configurations (\textsf{MD} \emph{frames}) of the shortest substate (2 ns) are taken as reference structures
for the anisotropic $\beta$-Gaussian network model (\textsf{$\beta$-GM}) \cite{Micheletti:2004:Proteins:15103627}.
This model retains only two centroids per amino acid: the $\mathrm{C}_\alpha$ atom
and a bead representing the sidechain. The free energy profile is approximated, for small deviations
from the reference structure, with a network of anisotropic springs of equal strength connecting all
the centroids within the cutoff distance of 7.5 \AA. Due to the quadratic nature of the interactions, the \textsf{DCM} $\mathcal{C}$
and the eigenspace $(\lambda_k,| \lambda_k \rangle)$ of a given protein structure are obtained inverting the resulting effective free energy matrix $\mathcal{M}$ (eq. 1-7 of \cite{Micheletti:2004:Proteins:15103627}) as
$\mathcal{C}= \mathcal{M}^{-1}$, where $\mathcal{C} | \lambda_k \rangle = \lambda_k | \lambda_k \rangle,\quad (\lambda_k > \lambda_{k+1})$.



Note that the total number of Degrees of Freedom (\textsf{DoF}) is $3 N - 6 = 636$, since $\mathcal{M}$ always presents six null modes (corresponding to global symmetries
of the model): the matrix inversion is obviously meaningful only within the subspace orthogonal to $\mathrm{ker}(\mathcal{M})$
\cite{Micheletti:2004:Proteins:15103627}.
The \emph{local} spectral statistics (probing correlations on a scale comparable with the mean level spacing) of the obtained ensemble of \textsf{DCM} is then compared with the predictions of
random correlation matrices. Analogous results (not included here), have been obtained on other
substates of
the same trajectory and on a MD simulation of G protein \cite{Pontiggia:2007:}, a
56 residues long protein, as a preliminary validation of the conclusions
drawn
in the present work on different case studies.
 
{\it Random Matrix Predictions.}
The Wishart-Laguerre (\textsf{WL}) \cite{mehta,wishart} ensemble of random covariances includes $N\times N$ matrices $\mathcal{W}$ of the form
$ \mathcal{W} = (1/T)\mathcal{X} \mathcal{X}^{\mathrm{t}}$, 
where $\mathcal{X} $ is an $N\times T$ matrix containing $N$ time series
of $T$ independent elements drawn from a Gaussian distribution with zero mean and \emph{fixed} variance, $\mathcal{X} \sim\mathrm{N}(0,\sigma^2)$.
Since $\mathcal{W}$  is the covariance matrix of a maximally random
data set, it is usually the optimal candidate as a `null model' with the lowest degree of built-in information
to compare empirical data with. This program has been implemented 
on financial data \cite{laloux},
internet routers networks \cite{bart}, EEG data \cite{seba} and atmospheric correlations \cite{sant} among others. 
The requirement of fixed data variance $\sigma^2$ guarantees rotational invariance and thus the exact solvability of \textsf{WL}.
Nevertheless, it makes the comparison inappropriate in the present case, where the \emph{empirical} `data matrix' $\tilde{\mathcal{X}}$ is not accessible
and its row variances $\tilde{\sigma}^2_i$ (corresponding to the $\mathcal{C}_{ii}$ entries) turn out to be unevenly spread.
We thus consider a slightly improved \textsf{$\sigma$-WL} model (non-invariant) where different variances
are randomly assigned to each data row. Details about this improved null model, unimportant for the present discussion, will be presented elsewhere; it suffices to say that the individual statistics is rather insensitive to the distribution of variances allocated to the rows.

As a typical example of local level statistics, we mainly consider 
the Individual Eigenvalue Spacing (\textsf{IES}) $s_k$ defined as \cite{muller} $s_k=(\lambda_{k}-\lambda_{k+1})/\langle\lambda_{k}-\lambda_{k+1}\rangle$,
where the average $\langle\cdot\rangle$ is taken over many samples and clearly $\langle s_k\rangle=1$ for any $k$. 
We numerically found that for \textsf{$\sigma$-WL} the \textsf{IES} is well approximated by
the Brody \cite{brody1973} one-parameter distribution, $p_\beta(s) = c_\beta (1 + \beta) s^\beta \exp(- c_\beta s^{1 + \beta})$, with $c_\beta = [\Gamma((2 + \beta)/(1 + \beta))]^{1 + \beta}$, and $\beta\approx 0.84 \pm 0.02$ (obtained from a one-parameter fit). Analytical predictions for the level distributions in non-invariant ensembles are generally lacking. The Brody distribution is the simplest 
and most commonly employed fitting formula for non-invariant \textsf{RMT} spacings (see e.g. \cite{lecaer} and references therein), interpolating between the limiting Poisson $(\beta=0)$ and Wigner $(\beta=1)$ distributions.


{\it Results and Discussion.}
The \textsf{RMT} local statistics is used here as a null model for two sets of spectra: $\{\lambda_k^{(j)}\}$ (the $k$-th bare eigenvalue
of $j$-th \textsf{DCM} sample, $k=1$ being the largest) and $\{\mu_k^{(j)}:= \lambda_k^{(j)} (3 N - 6)/\mathrm{Tr}[\mathcal{C}^{(j)}] \}$. The $\mu$'s
are normalized so that their sum reproduces the number of \textsf{DoF}. The $k$-th eigenvector triples $\{\mathsf{v}_{k,x}^{(j)},\mathsf{v}_{k,y}^{(j)},\mathsf{v}_{k,z}^{(j)}\}$ are also extracted.





\begin{figure}[!tpb]
 \centerline{\includegraphics[width=7cm]{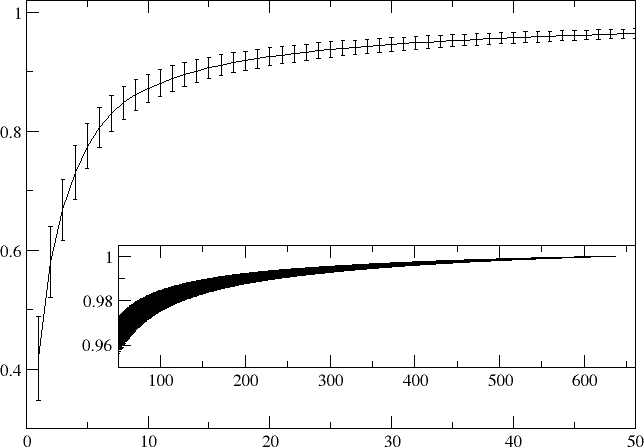}}
\caption{fraction of the protein's fluctuation $f_n$ (eq. \eqref{fn}) as a function of the first $n$ eigenvalues; the error bars are calculated as standard deviations.}
\label{fraction}
\end{figure}

The first quantity we analyzed is the average partial trace, or cumulative fraction of captured motion, defined as:

\begin{eqnarray}\label{fn}
f_n = \Big\langle\frac{1}{\mathrm{Tr}[\mathcal{C}]} \sum_{k=1}^n\ \lambda_k\Big\rangle \equiv \frac{1}{3 N - 6} \Big\langle\sum_{k=1}^n\ \mu_k\Big\rangle
\end{eqnarray}
and plotted in fig. \ref{fraction}.
As expected, the first 3-4 eigenvalues capture more than $70\%$ of the protein's overall mobility, and this value is typically larger in \textsf{MD} simulations \citep{amadei93}.
The very narrow dispersion validates in a statistical sense the persistence
of this feature during the \textsf{MD} simulation.

\begin{figure}[!tpb]
 \centerline{\includegraphics[width=7cm]{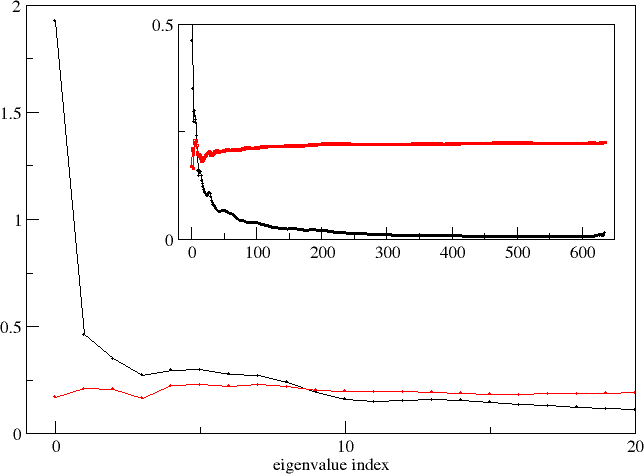}}
\caption{relative dispersion of eigenvalues.}
\label{dispersion}
\end{figure}
In order to statistically characterize each eigenvalue we plot in fig. \ref{dispersion} its relative dispersion (stdev/mean) {\it vs.} its index $k$: 
a low ratio signals a strong localization. The $\mu$'s display an almost constant ratio, suggesting a certain stability 
in the distribution of the fraction of total mobility captured by each mode. In comparison, the bare $\lambda$'s ratio rapidly decays to very low
values after crossing the range spanned by the $\mu$'s approximately between the 3rd and the 4th eigenvalue, indicating a broad dispersion
of the largest eigenvalues.

\begin{figure}[!tpb] 
 \centerline{\includegraphics[width=10.5cm]{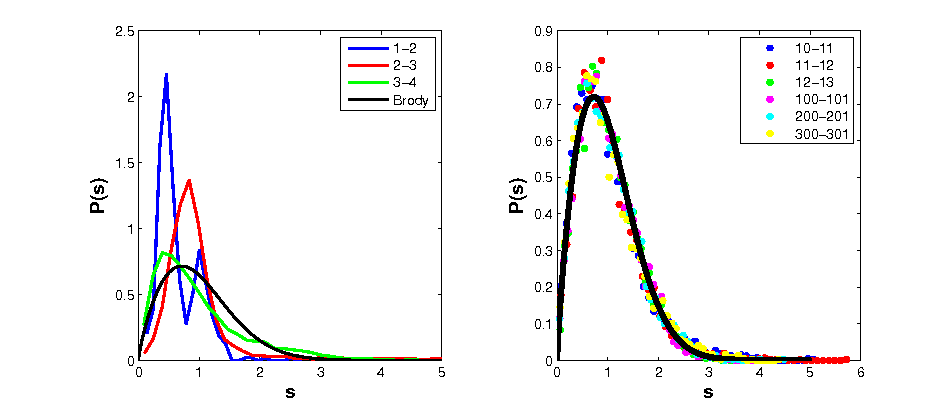}}
\caption{left, level spacing distributions of the first $3$ $\lambda_k$; right, samples of $\lambda_k$ spacings from the bulk.}
\label{spacing_lambda_1}
\end{figure}

\begin{figure}[!tpb]
 \centerline{\includegraphics[width=10.5cm]{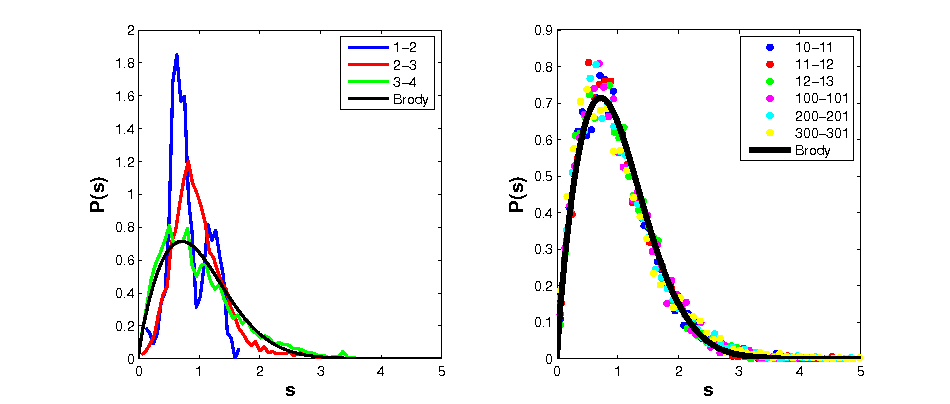}}
\caption{left, level spacing distributions of the first $3$ $\mu_k$; right, samples of $\mu_k$ spacings from the bulk.}
\label{spacing_mu_1}
\end{figure}

In contrast with the different dispersion properties of absolute and normalized eigenvalues, the spacing distributions in the bulk show a remarkable universal pattern. In figs. \ref{spacing_lambda_1} and \ref{spacing_mu_1}, the distributions of the $\lambda$'s and $\mu$'s are
fitted with a Brody distribution. 
With a $\chi^2$ test, the Brody hypothesis can be rejected with high confidence ($1\%$ level) for the first 3 spacings in both cases, while the subsequent ones give overall quite a good agreement with a fit parameter $\beta = 0.8 \pm 0.1$
(standard error among the first 100 spacings).
The same $\beta$ (within the statistical bounds) fits the spacing distribution for the null \textsf{$\sigma$-WL} model, indicating a strong degree of randomness in the largest fraction of the \textsf{DCM} spectra.

\begin{figure}[!tpb]
 \centerline{\includegraphics[width=9cm]{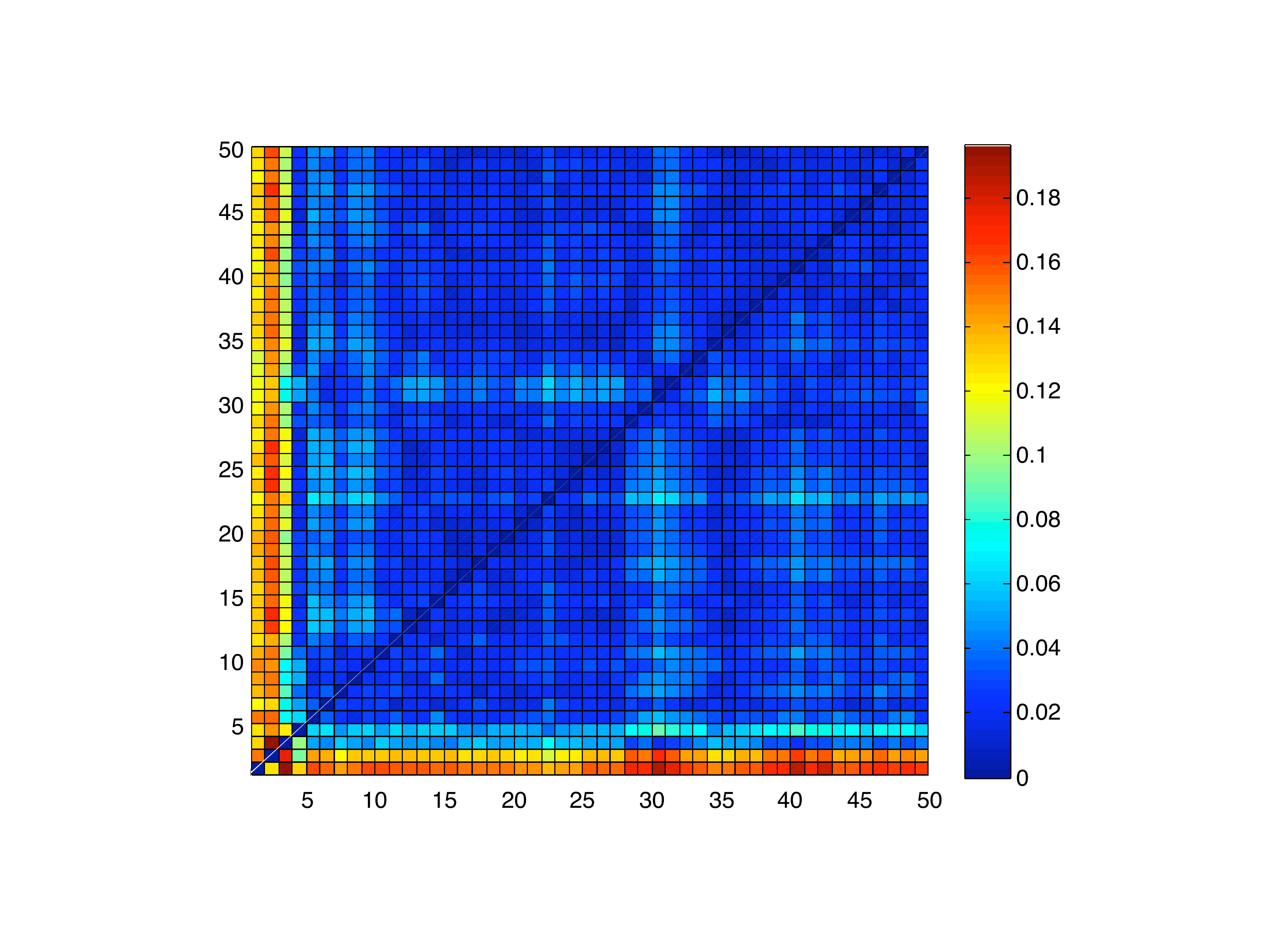}}
\caption{\textsf{KS} distances among spacing distributions of the $\lambda$ (top triangle) and $\mu$ (bottom triangle).}
\label{KS}
\end{figure}

The analysis of level spacings is completed with a Kolmogorov-Smirnov (\textsf{KS}) test among all pairs of spacing distributions. Fig. \ref{KS} shows the color-coded values of the \textsf{KS} distances
between the cumulative distributions. The same distribution appears to be closely shared by the spacings beyond the 4th, while no ``partner'' is found for the top four ones. Therefore, the sought significance criterion can be easily expressed as follows: \textit{there exists a transition between a few collective modes, characterized by a non-standard local level statistics, and the bulk of modes sharing the same quasi-universal distribution}.

\begin{figure}[!tpb]
 \centerline{\includegraphics[width=7cm]{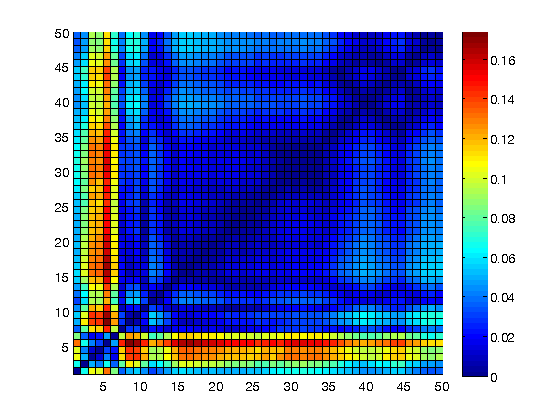}}
\caption{\textsf{KS} distances among distributions of the $\mathcal{C}$ eigenvectors.}
\label{vectors}
\end{figure}

\begin{figure}[!tpb]
 \centerline{\includegraphics[width=7cm]{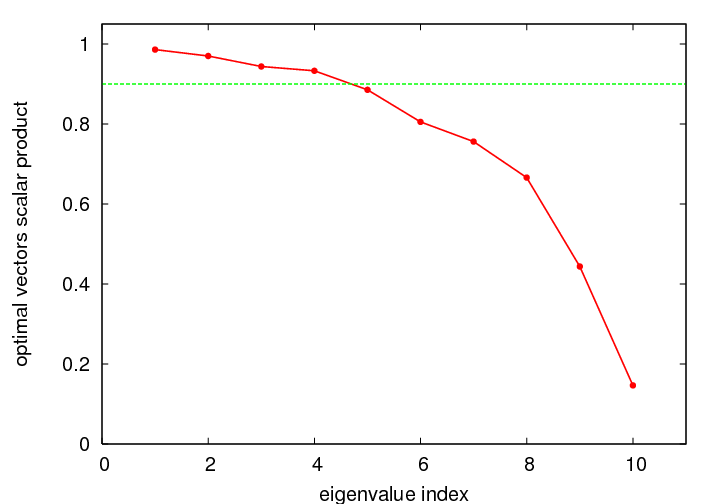}}
\caption{scalar products of optimal vectors pairs. The green line, indicating the 0.9 threshold, is shown as a guide to the eye.}
\label{optimal}
\end{figure}

In order to further validate these indications, in fig. \ref{vectors} we report a color-coded table of \textsf{KS} distances among eigenvector distributions. The entry $(\ell,k)$ 
represents the \textsf{KS} distance between the $\mathsf{v}_\ell$ and $\mathsf{v}_k$ cumulative distributions, after a proper weighting of the three Cartesian components.
Again, their distribution stabilizes only from the $5-6$th eigenvector onwards, while the first $4$ clearly display a very poor overlap
with the full set.

As a final check, we divided the \textsf{MD} trajectory
in two halves and computed the \textsf{DCM} for each sub-trajectory. We then applied the method
introduced in \cite{Pontiggia:2008:bj} to determine an optimal redefinition of the orthonormal basis vectors of the two essential spaces in order to quantify the degree of overlap between the two sets. Specifically, the redefined basis vectors in one set are ranked in order of decreasing overlap with the linear space spanned by the vectors in the other set.
The statistical significance of the first modes is confirmed by the high overlap of the first few optimal eigenvector pairs. This criterion, based on the properties of the essential space vectors rather than the eigenvalues, identifies about 4 relevant modes for the conserved subspace with a confidence level of about $90\%$ (see fig. \ref{optimal}).


{\it Summary and Outlook.}
In the present paper we applied \textsf{RMT} tools to an \textit{ensemble} of covariance matrices obtained in a biophysical context. Making use of an anisotropic \textsf{ENM}, applied to each configuration of a
\textsf{MD} simulation, we produced a large ensemble of covariance matrices for Adenylate Kinase.
The statistical properties of these \textsf{DCM} eigenspaces have been compared with universal \textsf{RMT} predictions, such as the Brody level spacing distribution of a suitable random ensemble.

The present study highlights signatures of bulk universality and random-like behavior shared by all but the first 3-4 eigenmodes of the
analyzed \textsf{DCM} ensemble. The consequence is a \textit{quantifiable} separation between the few most significant modes, characterized by their own peculiar statistics,
and the bulk of quasi-random ones. Likely implications include a more precise identification of the collective variables describing the large-scale,
functionally relevant fluctuations of biological molecules.

The marriage between \textsf{RMT} techniques and models of protein dynamics is expected to have a broad range of applications. Possible directions for 
future investigation include i) the temporal evolution of the significancy pattern during the protein's exploration of different substates; ii) the study
of global spectral properties (such as density and higher order correlation functions) of \textsf{DCM} within a single substate and along the full trajectory; 
iii) correlation structure among functional sub-units of a single protein, involving only a finite fraction of highly connected components; and finally
iv) independent validation of the method on other protein structures and using \textsf{MD} covariance matrices rather than \textsf{ENM} ones.


{\it Acknowledgments.} We warmly thank C. Micheletti, F. Pontiggia, Y. Gerelli and G. Akemann for helpful discussions and advices. We are indebted with G. Colombo for making the MD simulations performed in \cite{Pontiggia:2007:} available to us and with A. Liguori for a careful reading of the manuscript. FC acknowledges the grant 2007JHLPEZ (MIUR).

\bibliography{rmt_and_enmV16}

\end{document}